\documentstyle[12pt]{article} 
\newcommand{\dsl}[1]{\not \! #1} 
\newcommand{\tr}{{\rm tr}}

\title{Thirring Model as a Gauge Theory 
\thanks{To be published in Nuclear Physics B.} } 
\author{%
Kei-Ichi KONDO
    \thanks{e-mail: kondo@cuphd.nd.chiba-u.ac.jp}
\vspace{1.2em}\\
  Department of Physics, Faculty of Science \\
  $\&$ Graduate School of Science and Technology,\\
         Chiba University, Chiba 263, Japan
}
\date{CHIBA-EP-87-REV,
\\ February 1995,
\\ hep-th/9502070}
\begin{document}
\maketitle
\centerline{Abstract}
We give another reformulation of the Thirring model (with
four-fermion interaction of the current-current type) as a
gauge theory and identify it with a gauge-fixed version of
the corresponding gauge theory according to the
Batalin-Fradkin formalism.  Based on this formalism, we
study the chiral symmetry breaking of the
$D$-dimensional Thirring model ($2<D<4$) with $N$
flavors of 4-component fermions. By constructing the gauge
covariant effective potential for the chiral order
parameter, up to the leading order of $1/N$ expansion, we
show the existence of the second order chiral phase
transition and obtain explicitly the critical number of
flavors $N_c$ (resp. critical four-fermion coupling $G_c$)
as a function of the four-fermion coupling $G$ (resp. $N$),
below (resp. above) which the chiral symmetry is
spontaneously broken.
\newpage

\section{Introduction}
\par
Recently the dynamical mass generation in the $D$ $(2 < D
<4)$ dimensional Thirring model \cite{Thirring} has been
extensively studied by several authors
\cite{KK91,GMRS91,HP94,HLL94,ACCP94,RS94}.
A starting point of the analysis of the Thirring model is
to introduce the vector auxiliary field to linearize the
four-fermion interaction.
The original Thirring model has of course {\it no local
gauge invariance}.  After the introduction of the
auxiliary field the gauge invariance is still absent.
Indeed the Thirring model is apparently rewritten into the
massive vector theory with which the fermion couples
minimally.
However the results of these papers contradict with each
other and are rather confusing.  Contradictions among
these papers arise because of discrepancies among different
regularization schemes adopted when making the theory
finite.
Although they treated the vector auxiliary field as a gauge
field despite the absence of manifest gauge symmetry
\cite{GMRS91,HP94},   there is no
principle to determine which regularization is a privileged
one to be selected.  In this sense the issue of dynamical
mass generation strongly depends on the adopted
regularization.
\par
There is a possibility of forcing gauge invariance in
such a model after an auxiliary vector field is
introduced, by introducing a St\"uckelberg scalar field.
Quite recently, Itoh et al. \cite{IKSY94} have proposed to
maintain manifest gauge symmetry by reformulating the
Thirring model truly as a gauge theory.  This was done by
using the hidden local symmetry \cite{BKY} as a guiding
principle.
They argue that the existence of manifest gauge symmetry can
draw more definite result on the induced Chern-Simons term.
For example, the Vafa-Witten theorem \cite{VW84} which does
not rely on the specific regularization can be applied to
this problem due to the existence of the gauge symmetry as
in (2+1)-dimensional QED (QED3). Hence, for even number of
2-component fermions, parity is not spontaneously broken,
since the induced Chern-Simons term of each fermion species
can be arranged in pair of opposite sign to cancel each
other.  The parity breaking configuration is not
energetically unstable.
According to this fact, we assume in this paper no parity
violation for $N$ flavors of 4-component fermions or $2N$
(even) flavors of 2-component fermions and pay attention to
the chiral symmetry breaking.  Of course, whether this
strategy is possible or not depends on the adopted
regularization scheme.

\par
The Thirring model is rewritten as a gauge-fixed version of
a gauge theory by introducing the St\"uckelberg field
$\theta$ in addition to the auxiliary vector field
$A_\mu$ which is now identified with the gauge field.
This is a consequence of the general formalism for the
constrained system by Batalin and Fradkin \cite{BF86}.
This gives the general procedure by which the system with
the second class constraint is converted to that with the
first class one and the new field which
is necessary to complete this procedure is called the
Batalin-Fradkin field \cite{BF87}.  In the massive gauge
theory the Batalin-Fradkin field is nothing but the
well-known St\"uckelberg field as shown in \cite{FIK90}.
\par
In this paper, based on the Batalin-Fradkin formalism we
give another reformulation of the Thirring model as a gauge
theory and interpret it as the gauge-fixed version of the
gauge theory in section 2.
The final form of the gauge theory is the same as that
given in \cite{IKSY94}, as should be.
Based on this formalism, we study in section 5 the
spontaneous breakdown of the chiral symmetry through the
effective potential obtained in section 4 for the order
parameter of the chiral symmetry,  the chiral condensate
$\langle \bar \psi \psi \rangle$.
Up to the leading order of $1/N$ expansion where $N$ is the
fermion flavor, we obtain explicitly the critical
number of flavors $N_c$ (resp. critical four-fermion
coupling $G_c$) as a function of  $G$ (resp. $N$), below
(resp. above) which the chiral symmetry is spontaneously
broken.
In section 3, we study the behavior of the vacuum
polarization of the gauge boson propagator with respect to
the source term for the fermion mass, which is necessary
to obtain the effective potential.
\par

\section{Thirring model as a gauge theory}
\par
In this paper we consider the $D$-dimensional Thirring model
($2<D<4$).
The Lagrangian density of the Thirring model is given by
\begin{eqnarray}
 {\cal L}_{Th}
 = \bar \psi^a i \gamma^\mu \partial_\mu \psi^a
 - m \bar \psi^a \psi^a
 - {G \over 2N}(\bar \psi^a \gamma^\mu \psi^a)^2,
 \label{Thirring}
\end{eqnarray}
where $\psi^a$ is a 4-component Dirac spinor with an index
$a$ being summed over from 1 to $N$ and $\gamma_\mu
(\mu=0,1,2,...,D-1)$ are $4 \times 4$ gamma matrices
satisfying the Clifford algebra
$\{ \gamma_\mu, \gamma_\nu \} = 2 g_{\mu\nu} {\bf 1}$.
\par
By introducing an auxiliary vector field
$A_\mu$, the Lagrangian is rewritten as
\begin{eqnarray}
 {\cal L}_{Th'} = \bar \psi^a  i \gamma^\mu (\partial_\mu
 - {i \over \sqrt{N}} A_\mu) \psi^a
 - m \bar \psi^a \psi^a
 + {1 \over 2G} A_\mu^2.
 \label{th'}
\end{eqnarray}
Here $A_\mu$ denotes the massive vector boson which does
not have the corresponding kinetic term and the Lagrangian
Eq.~(\ref{th'}) has no gauge symmetry.
A crucial point in formulating the Thirring model as a
gauge theory based on the Batalin-Fradkin (BF) formalism
is the existence of the kinetic term for the
field $A_\mu$.
Such a viewpoint based on the hidden local symmetry has
been already emphasized in \cite{IKSY94}.
In the massive fermion case, the kinetic term is generated
through the radiative correction to the gauge boson
propagator.  Actually the massive Thirring model is mapped
into the equivalent gauge theory by bosonization,
especially into the Maxwell-Chern-Simons theory in three
dimensions, as discussed in a subsequent paper
\cite{Kondo95b} where
an advantage of keeping the manifest gauge invariance will
be elucidated in the intermediate step of bosonization.
For the massless Thirring model, on the other hand, this
problem is somewhat subtle
\cite{Redlich84}.  However, even in the massless Thirring
model, such a kinetic term which is signaled by the
appearance of a pole in the gauge boson propagator is
generated through the dynamical generation of the fermion
mass $m_f$, which was shown in \cite{IKSY94} at one-loop
level.  Moreover the bosonization of the massless
Thirring model has been discussed in \cite{IKSY94}.
For a while, we assume the existence of such a kinetic
term.  This problem is again discussed in section 5.2.
\par
By making use of the St\"uckelberg field  $\theta$ which is
shown \cite{FIK90} to be identified with the BF field
\cite{BF87} in the general formalism for the constrained
system \cite{BF86},  the original Thirring model is
identified with the gauge-fixed version of the gauge
theory with the Lagrangian \cite{KK91}:
\begin{eqnarray}
 {\cal L}_{Th''}
 = \bar \psi^a i \gamma^\mu (\partial_\mu
 - {i \over \sqrt{N}} A_\mu) \psi^a
 - m \bar \psi^a \psi^a
 + {1 \over 2G}(A_\mu - \sqrt{N}\partial_\mu \theta)^2,
\label{th''}
\end{eqnarray}
which possesses a $U(1)$ gauge symmetry and is
invariant under the transformation:
\begin{eqnarray}
 \psi_a \longmapsto \psi_a ' = e^{i\phi} \psi_a,
 A_\mu \longmapsto
 A_\mu ' = A_\mu + \sqrt{N} \partial_\mu \phi,
 \theta \longmapsto \theta ' = \theta +  \phi.
\end{eqnarray}
Actually, if we take the unitary gauge
$\theta'=0$ $(\phi = - \theta)$, then the
Lagrangian~(\ref{th''}) reduces to the
Lagrangian~(\ref{th'}), as pointed out in \cite{KK91}.
The $A_\mu$ in Eq.~(\ref{th''}) is regarded with the
massless gauge field, in sharp contrast to the massive
vector boson
$A_\mu$ in Eq.~(\ref{th'}).
Advantages of the existence of such a gauge symmetry in
the analysis of the Thirring model are emphasized recently
in \cite{IKSY94}.
\par
BF formalism starts from the Becchi-Rouet-Stora (BRS)
invariant formulation. It is well known that the total
Lagrangian which is invariant under the BRS transformation
is obtained by adding the gauge-fixing term and the
Faddeev-Popov (FP) ghost term
${\cal L}_{GF+FP}$ to the Lagrangian ${\cal L}_{Th''}$:
\begin{eqnarray}
 {\cal L}_{GF+FP}
 = -i \delta_B(\bar c f[A,c,\bar c,B,\theta]),
\end{eqnarray}
which is itself BRS-invariant because of nilpotency
$\delta_B^2 * = 0$.
Such a BRS transformation is given by
\begin{eqnarray}
 \delta_B A_\mu(x) &=&  \partial_\mu c(x),
\nonumber\\
 \delta_B B(x) &=& 0,
\nonumber\\
 \delta_B c(x) &=& 0,
\nonumber\\
 \delta_B \bar c(x) &=&  i B(x),
\nonumber\\
 \delta_B \theta(x) &=& {1 \over \sqrt{N}} c(x) ,
\nonumber\\
 \delta_B \psi^j(x) &=&  {i \over \sqrt{N}}  c(x) \psi^j(x),
\end{eqnarray}
where $c(x)$ and $\bar c(x)$ are FP ghost fields, and
$B(x)$ is the Nakanishi-Lautrap Lagrange multiplier field.
For the choice:
$
f[A,c,\bar c,B,\theta]=F[A,\theta]+{\xi \over2} B,
$
we obtain
\begin{eqnarray}
 {\cal L}_{GF+FP}
 = B F[A,\theta] + {\xi \over 2} B^2
 + i \bar c \left(
 {\delta F[A,\theta] \over \delta A_\mu} \partial_\mu
 + {1 \over \sqrt{N}}
 {\delta F[A,\theta] \over \delta \theta} \right) c .
\end{eqnarray}
When $F[A,\theta]$ is linear separately in $A_\mu$ and
$\theta$, FP ghost decouples completely from the system.
Then, after performing the integration over $B$, we obtain
the gauge-fixing term ${\cal L}_{GF}$:
\begin{eqnarray}
 {\cal L}_{GF} = - {1 \over 2\xi}(F[A,\theta])^2,
\end{eqnarray}
with a gauge-fixing parameter $\xi$.
In the covariant gauge given by
$F[A,\theta]=\partial^\mu A_\mu$, the BF field $\theta$ is
not decoupled except the Landau gauge $\xi=0$
\cite{KK91,IKSY94}.

\par
By choosing the $R_\xi$ gauge,
$F[A,\theta]=\partial^\mu A_\mu + \sqrt{N} {\xi \over G}
\theta$, the BF field $\theta$ is completely
decoupled independently of $\xi$ and the total Lagrangian
${\cal L}_{Th'''} = {\cal L}_{Th''}+{\cal L}_{GF}$ reduces
to the following form \cite{IKSY94}:
\begin{eqnarray}
 {\cal L}_{Th'''}
 &=& {\cal L}_{\psi, A} + {\cal L}_\theta,
\nonumber\\
 {\cal L}_{\psi, A}
 &=& \bar \psi^a i \gamma^\mu
 (\partial_\mu - {i \over \sqrt{N}}A_\mu) \psi^a
 - J \bar \psi^a \psi^a
 + {M^2 \over 2} A_\mu^2
 - {1 \over 2\xi} (\partial^\mu A_\mu)^2,
\nonumber\\
 {\cal L}_\theta
 &=& {1 \over 2} (\partial_\mu \theta)^2
 - {\xi \over 2G} \theta^2,
 \quad M^2 \equiv {1 \over G},
 \label{Th'''}
\end{eqnarray}
where we have introduced an {\it infinitesimal} external
source $J(>0)$ for the fermion mass to study the spontaneous
chiral symmetry breaking and $J$ is eventually adjusted to
go to zero.
\par
As pointed out in \cite{IKSY94},
the existence of such a gauge symmetry enables us to apply
the Vafa-Witten theorem \cite{VW84} in the same way as in
the three-dimensional gauge theory, for example, QED3.
According to Vafa-Witten \cite{VW84}, energetically
favorable is a parity conserving configuration: all the
2-component fermions have the same absolute value and
half of them acquires positive masses and the other half
negative masses.
Moreover, the parity violating pieces including the induced
Chern-Simons term don't appear in the gauge sector whenever
the number of 2-component fermion is even, in agreement
with various analyses
\cite{ABKW86b,Poly88,RY86,CCW91,KEIT95}.
Therefore we consider the pattern of symmetry breaking not
for the parity but the chiral symmetry.  In this paper we
investigate the chiral condensate
$\langle \bar \psi^a \psi^a \rangle$
as an order parameter for the chiral symmetry breaking.
The chiral symmetry is defined for the 4-component fermion
by make use of a $4 \times 4$ matrix $\gamma_D$, which
anticommutes with all the gamma matrices, see
\cite{ABKW86}.

\section{Vacuum polarization}
\par
Before beginning the calculation, it is instructive to give
some comments on the choice of regularizations.  The
regularization must be chosen in such a way to preserve the
gauge symmetry.  There are various gauge-invariant
regularization methods to calculate the vacuum
polarization.  For example,
1) Pauli-Villars
\cite{DJT81,Redlich84,KS85,Alvarez90},
2) lattice \cite{So85,CL89},
3) analytic \cite{PST92},
4) dimensional \cite{Martin90,DW93},
5) Zavialov class \cite{Novotny92},
6) parity-invariant Pauli-Villars (variant of chiral gauge
invariant Pauli-Villars by Frolov and Slavnov)
\cite{Kimura94},
7) high covariant derivative \cite{AF90},
8) zeta-function \cite{GMSS85}.
\par
In the case of odd number of 2-component fermions,
however, a peculiarity arises in (2+1) dimensions
where the parity-violating Chern-Simons term is induced
through fermion loop correction even if the bare Lagrangian
does not contain such a term.
However the coefficient of such an induced Chern-Simons term
depends upon how to regularize the ultraviolet divergences.
The ordinary Pauli-Villars regularization (or lattice
regularization) in (2+1) dimensions explicitly breaks
parity invariance due to the regulator fermion mass (resp.
the Wilson term) and this parity violating effect remains
finite even after the regulator is removed, i.e., the
regulator masses (resp. lattice cutoff) tend to infinity
(resp. zero).  Hence the parity violating Chern-Simons
term arises even in the symmetric phase where the fermion
mass is not dynamically generated.
Nevertheless it is shown \cite{Kimura94} that one can
develop the parity-invariant Pauli-Villars regularization
method in which the regularization procedure by itself does
not induce any Chern-Simons term and that this
parity-invariant Pauli-Villars regularization gives at
least to the one-loop level the same result as that in
dimensional and analytic regularizations.
For even number of flavors, if one adopts a Pauli-Villars
regularization scheme and chooses the masses of the
regularizations with alternate signs, no parity breaking
arises.    If instead one uses zeta-function
regularization, there is always parity breaking
\cite{GMSS85} irrespective of the number of flavors.
\footnote{The author thanks the referee for
informing this result.}

\par
In what follows we follow the Euclidean formulation.
By making use of a gauge-invariant Pauli-Villars
regularization \cite{IZ80}, it is shown
\cite{Hands94}  that the one-loop vacuum polarization
tensor has the form:
\begin{eqnarray}
 \Pi_{\mu\nu}^{(1)}(k)
 = \left( \delta_{\mu\nu}-{k_\mu k_\nu \over k^2} \right)
  \Pi_T^{(1)}(k^2;J),
\end{eqnarray}
and
\begin{eqnarray}
 \Pi_T^{(1)}(k^2;J)
 = - k^2 {2\tr(1) \Gamma(2-D/2) \over (4\pi)^{D/2}}
\int_0^1 d\alpha {\alpha(1-\alpha) \over
[\alpha(1-\alpha)k^2+J^2]^{2-D/2}},
\end{eqnarray}
which is rewritten into a compact form:
\begin{eqnarray}
 \Pi_T^{(1)}(k^2;J)
 = -  {\tr(1) \Gamma(2-D/2) \over 3(4\pi)^{D/2}}
{k^2 \over J^{4-D}}
{}_2F_1(2,2-{D \over 2},{5 \over 2};-{k^2 \over 4J^2}),
\end{eqnarray}
with ${}_2F_1(a,b,c;z)$ being a hypergeometric function.

{}From the mathematical identity for the hypergeometric
function,
\begin{eqnarray}
{}_2F_1(a,b,c;z)
&=&
{\Gamma(c)\Gamma(b-a) \over \Gamma(b)\Gamma(c-a)}
(-z)^{-a} {}_2F_1(a,1-c+a,1-b+a,z^{-1})
\nonumber\\&&
+ {\Gamma(c)\Gamma(a-b) \over \Gamma(a)\Gamma(c-b)}
(-z)^{-b} {}_2F_1(b,1-c+b,1-a+b,z^{-1}),
\end{eqnarray}
the vacuum polarization function is rewritten as
\begin{eqnarray}
&&  \Pi_T^{(1)}(k^2;J) /\tr({\bf 1})
\nonumber\\ &=&  -4
{\sqrt{\pi} \Gamma(D/2)\Gamma(2-D/2) \over
4^{D/2}(4\pi)^{D/2}\Gamma(1/2+D/2)} k^{D-2}
{}_2F_1(2-{D \over 2},{1 \over 2}-{D \over 2},1-{D \over
2};-{4J^2 \over k^2})
\nonumber\\&&
-4  {\Gamma(-D/2) \over (4\pi)^{D/2}}{J^D \over k^2}
{}_2F_1(2,{1 \over 2},1+{D \over 2};-{4J^2 \over k^2}).
\end{eqnarray}
Taking into account the power-series expansion of the
hypergeometric function,
\begin{eqnarray}
&&
{}_2F_1(2-{D \over 2},{1 \over 2}-{D \over 2},
1-{D \over 2};-{4J^2 \over k^2})
\nonumber\\
&=&
 1 - {2(4-D)(D-1) \over D-2}{J^2 \over k^2}
+ {2(4-D)(6-D)(D-1)(3-D) \over (D-2)(4-D)}{J^4 \over k^4}
\nonumber\\ &&
+ {\cal O}(J^6/k^6),
\end{eqnarray}
we can show
\begin{eqnarray}
 \Pi_T^{(1)}(k^2;J) = - f_0(k) + J^2 f_2(k) + J^D f_D(k)
 + J^4 f_4(k) + {\cal O}(J^{D+2}, J^6),
\label{vpexpansion}
\end{eqnarray}
where
\begin{eqnarray}
 f_0(k) &=& 4
{\tr(1)\sqrt{\pi} \Gamma(D/2)\Gamma(2-D/2) \over
4^{D}\pi^{D/2}\Gamma(1/2+D/2)} k^{D-2}
\equiv  r_D k^{D-2} >0,
 \\
 f_2(k) &=&  r_D {2(4-D)(D-1) \over D-2} k^{D-4} >0,
 \\
 f_D(k) &=&
 -4  {\tr(1)\Gamma(-D/2) \over (4\pi)^{D/2}}{1 \over
k^2}<0,
 \\
 f_4(k) &=& -  r_D {2(4-D)(6-D)(D-1)(3-D) \over
(D-2)(4-D)} k^{D-6} .
\end{eqnarray}
Here note that $f_0$ and $f_2$ are positive functions and
$f_D$ is negative one for $2<D<4$, since
$\Gamma(x)>0$ for $x>0$ and $-1>x>-2$.
\par
Rigorously speaking, the expression
Eq.~(\ref{vpexpansion}) is meaningful only when $|{2J \over
k}|<1$. On the other hand, in the region
$|{k \over 2J}|<1$, we must use
\begin{eqnarray}
 \Pi_T^{(1)}(k^2;J)  &=& -  {\tr(1) \Gamma(2-D/2) \over
3(4\pi)^{D/2}} k^2 \Big[J^{D-4} - {2-D/2 \over 5} k^2
J^{D-6}
\nonumber\\&&
+ {\cal O}(k^4 J^{D-8}) \Big].
\end{eqnarray}
However inclusion of this contribution does not change the
result at all. This is shown in subsection 5.4.

\section{Effective potential by Inversion}
The chiral order parameter $\phi$ is obtained in the scheme
of $1/N$ expansion as
\begin{eqnarray}
\phi &:=& \langle \bar \psi \psi \rangle
= {\partial \over \partial J}
\left[ \ln \det (i\gamma^\mu \partial_\mu + J)
- {1 \over 2} \ln \det [D_{\mu\nu}^{(1)}]^{-1} \right]
+ {\cal O}(1/N^2),
\label{order}
\end{eqnarray}
where $D_{\mu\nu}^{(1)}$ is the leading $1/N$ gauge boson
propagator \cite{SW92}.
Taking into account the relation:
\begin{eqnarray}
[D_{\mu\nu}^{(1)}(k)]^{-1} = [D_{\mu\nu}^{(0)}(k)]^{-1}  +
{1 \over N} \Pi_{\mu\nu}^{(1)}(k) ,
\end{eqnarray}
with $D_{\mu\nu}^{(0)}$ being the free photon propagator,
it is not difficult to show \cite{Kondo95} that the equation
Eq.~(\ref{order}) is rewritten as
\begin{eqnarray}
\phi
= \langle \bar \psi \psi \rangle_0
+ {1 \over 2N} \int {d^Dk \over (2\pi)^D}
D_{\mu\nu}^{(1)}(k)
{\partial \over \partial J} \Pi_{\mu\nu}^{(1)}(k)
 + {\cal O}(1/N^2),
 \label{op}
\end{eqnarray}
where
$\langle \bar \psi \psi \rangle_0$
is the leading part defined by
\begin{eqnarray}
 \langle \bar \psi \psi \rangle_0
 = \int {d^Dp \over (2\pi)^D} \tr {1 \over {\dsl p}+J},
\end{eqnarray}
and $D_{\mu\nu}^{(1)}$ is
derived from the Lagrangian~(\ref{Th'''}):
\begin{eqnarray}
 D_{\mu\nu}^{(1)}(k)
 = {1 \over M^2-\Pi_T^{(1)}(k^2)}
 \left( \delta_{\mu\nu}-{k_\mu k_\nu \over k^2} \right)
  +  {\xi \over k^2+\xi M^2}{k_\mu k_\nu \over k^2},
  M^2 = G^{-1}.
  \label{propagator}
\end{eqnarray}
Here the limit $\xi \rightarrow \infty$ corresponds to the
unitary gauge and $\xi = 0$ to the Landau gauge. In the
unitary gauge, it is impossible to take the $G \rightarrow
\infty$ limit in $D_{\mu\nu}^{(1)}(k)$.
A merit of the expression Eq.~(\ref{op})  is that
the chiral order parameter evaluated according to
Eq.~(\ref{op}) gives  the {\it gauge-covariant}, i.e.,
gauge-parameter-independent result:
\begin{eqnarray}
D_{\mu\nu}^{(1)}(k) {\partial \over \partial J}
\Pi_{\mu\nu}^{(1)}(k)
= (D-1) [M^2-\Pi_T^{(1)}(k^2;J)]^{-1}
{\partial \over \partial J} \Pi_T^{(1)}(k^2;J),
\label{contract}
\end{eqnarray}
since
$
\Pi_{\mu\nu}^{(1)}(k)
= \left( \delta_{\mu\nu}-{k_\mu k_\nu \over k^2} \right)
\Pi_T^{(1)}(k^2;J)
$
is transverse owing to the gauge invariance.  Therefore all
the following results are independent of the choice of the
gauge parameter $\xi$.
\par
First of all,
by introducing an ultraviolet (UV) cutoff $\Lambda_f$, the
leading part reads
\begin{eqnarray}
 \langle \bar \psi \psi \rangle_0/\Lambda_f^{D-1}
 &=& C_D \tr(1) J \int_0^{\Lambda_f} dp
 {p^{D-1} \over p^2+J^2}
\nonumber\\
 &=& C_D \tr(1)  D^{-1}
{\Lambda_f \over  J}
{}_2F_1 (1,{D \over 2},
     1 + {D \over 2},-{\Lambda_f^2 \over J^2}) ,
 \label{}
\end{eqnarray}
with
\begin{eqnarray}
C_D := {1 \over 2^{D-1}\pi^{D/2}\Gamma(D/2)}.
\end{eqnarray}
This gives the expansion:
\begin{eqnarray}
 \langle \bar \psi \psi \rangle_0/\Lambda_f^{D-1}
&=& C_D \tr(1)  D^{-1}
 \Gamma (1 + {D \over 2})
  \Biggr[  {\Gamma (-1 + {D \over 2}) \over
  \Gamma ({ D \over 2})^2}  {J \over \Lambda_f}
\nonumber\\ &&
 +  {J^{D-1} \over \Lambda^{D-1}}
    \Gamma (1 - {D \over 2})
 -      { \left( 1 - {D \over 2} \right)
   \Gamma (-1 + {D \over 2})  \over
        \left( 2 - {D \over 2} \right)
 \Gamma ({D \over 2})^2 }  {J^{3} \over \Lambda_f^{3}}
 +       {\rm O}(J^5)  \Biggr] .
 \label{opfree}
\end{eqnarray}
\par
Next, substituting Eq.~({\ref{vpexpansion}) into
Eq.~(\ref{contract}),
we obtain
\begin{eqnarray}
&&
D_{\mu\nu}^{(1)}(k) {\partial \over \partial J}
\Pi_{\mu\nu}^{(1)}(k)
\nonumber\\
&=& (D-1)[M^2+f_0(k)]^{-1} [2 f_2(k)J  + Df_D(k) J^{D-1}]
\nonumber\\ &&
 + (D-1)
 [M^2+f_0(k)]^{-2}\{2f_2^2(k)+4 f_4(k)[M^2+f_0(k)]\}J^3
\nonumber\\ &&
+ {\cal O}(J^{D+1},J^{2D-1},J^5).
\end{eqnarray}
Therefore we conclude
\begin{eqnarray}
&&
{1 \over 2} \int {d^Dk \over (2\pi)^D}
D_{\mu\nu}^{(1)}(k) {\partial \over \partial J}
\Pi_{\mu\nu}^{(1)}(k)
\nonumber\\ &=&  K_1 J  - P_1 J^{D-1}  + Q_1 J^3
 + {\cal O}(J^{D+1},J^{2D-1},J^5),
\end{eqnarray}
where
\begin{eqnarray}
K_1 &=&  {D-1 \over 2}C_D  \int_0^{\Lambda_p} k^{D-1} dk
 {2 f_2(k) \over [M^2+f_0(k)]} >0,
 \label{K1}
\\
P_1 &=& - {D-1 \over 2}C_D  \int_0^{\Lambda_p} k^{D-1} dk
 {Df_D(k) \over [M^2+f_0(k)]} >0.
\end{eqnarray}
Here we have introduced another UV cutoff $\Lambda_p$ for
the gauge-field momentum and note that $K_1$ and $P_1$ are
positive.

\par
Thus, the chiral order parameter $\phi$ shows the following
dependence on the source term
$J$ for $2<D<4$:
\begin{eqnarray}
\phi := \langle \bar \psi \psi \rangle
 = K J - P J^{D-1} - Q J^3
 + {\cal O}(J^{D+1}, J^{2D-1}, J^5) ,
\end{eqnarray}
where the coefficients $K, P$ and $Q$ are given in the
form of power-series in $1/N$:
$K=K_0+K_1/N+{\cal O}(1/N^2)$ and so on. In particular, we
find from Eq.~(\ref{opfree}) and  Eq.~(\ref{K1}):
\begin{eqnarray}
 K_0 &=& {C_D \tr(1) \over D-2} \Lambda_f^{D-2} > 0,
\nonumber\\
 K_1 &=&  C_D r_D {2(4-D)(D-1)^2 \over D-2}
 \int_0^{\Lambda_p} dk
 {k^{2D-5} \over M^2+ r_D k^{D-2}} > 0,
\end{eqnarray}
and
\begin{eqnarray}
 P_0 =  - C_D \tr(1) D^{-1} \Gamma(1+{D \over 2})
\Gamma(1-{D \over 2}) > 0 .
\end{eqnarray}
\par
We can define the dimensionless order parameter $\varphi$
from $\phi$ by using a certain dimensionful quantity
$\alpha$ with the same dimension as the mass.  Here
$\alpha$ may be identified with the ultraviolet cutoff
$\Lambda$, the dynamically generated fermion mass
$m_f$ or defined from the dimensionful coupling constant
$G$.  Therefore we obtain
\begin{eqnarray}
 \varphi
 = {\langle \bar \psi \psi \rangle \over \alpha^{D-1}}
 = \tilde K \tilde J
 -  \tilde P \tilde J^{D-1}
 - \tilde Q \tilde J^3
 + {\cal O}(J^{D+1}, J^{2D-1}, J^5),
 \label{phi}
\end{eqnarray}
where the dimensionless coefficients are defined:
\begin{eqnarray}
 \tilde K := {K \over \alpha^{D-2}},
\quad
 \tilde Q := {Q \over \alpha^{D-4}},
\quad
 \tilde P = P,
\end{eqnarray}
as well as the dimensionless source:
\begin{eqnarray}
 \tilde J := {J \over \alpha}.
\end{eqnarray}

\par
Instead of taking the Legendre transform, we here adopt
the inversion method \cite{Fukuda}.
By inverting the equation~(\ref{phi}) in
terms of the source $J$, we obtain
\begin{eqnarray}
 \tilde J
 = \tau \varphi + B \varphi^{D-1} + A \varphi^3
 + {\cal O}(\varphi^{D+1}, \varphi^{2D-1}, \varphi^5).
 \label{inv}
\end{eqnarray}
For the inverted series Eq.~(\ref{inv}) to be consistent
with the original series Eq.~(\ref{phi}), the
coefficients in Eq.~(\ref{inv}) are determined:
\begin{eqnarray}
 \tau &=& \tilde K^{-1}
 =  \tilde K_0 - \tilde K_1/N + {\cal O}(1/N^2) ,
\nonumber\\
 B &=&  \tilde P \tilde K^{-1} \tau^{D-1}
 = \tilde P \tilde K^{-D} .
\end{eqnarray}
\par
The effective potential for the translation-invariant
expectation value
$\phi = \langle \bar \psi \psi \rangle$
(order parameter) is obtained from
the effective action $\Gamma[\phi]$ through the relation:
\begin{eqnarray}
  J = {\partial \over \partial \phi} V(\phi),
  V(\phi) := \Gamma[\phi]/\int d^Dx .
\end{eqnarray}
Then, for $2<D<4$, the effective potential for $\varphi$ is
obtained:
\begin{eqnarray}
  V(\varphi) = \alpha^{-D} V(\phi)
  = {\tau \over 2} \varphi^2
 + {B \over D} \varphi^{D} + {A \over 4} \varphi^4
 + {\cal O}(\varphi^{D+2}, \varphi^{2D}, \varphi^6).
 \label{EP}
\end{eqnarray}
The spontaneous breakdown of the chiral symmetry occurs if
the equation
$
{\partial \over \partial \phi} V(\phi) = J
$
has a non-trivial solution $\phi \not= 0$ even in the limit
$J \rightarrow 0$.
The most energetically favorable configuration is realized
at the absolute minima of the effective potential among
the stationary points.
As long as $B>0$, the phase
transition occurs at
$\tau=0$ and $\varphi$ has a nonzero value,
\begin{eqnarray}
\varphi \sim (-\tau/B)^{1/(D-2)},
\end{eqnarray}
for $\tau<0$.
Here we have neglected higher powers of $\varphi$, since
we are interested only in the neighborhood of the phase
transition point.  It is easy to see that the chiral phase
transition described by the effective potential
Eq.~(\ref{EP}) is the second order.  Up to the leading
order of $1/N$, there exists a critical number of flavors
which is given by
\begin{eqnarray}
 N_c = \tilde K_1/\tilde K_0 = K_1/K_0.
\end{eqnarray}
This shows that the critical number of flavors $N_c$ does
not depend on what quantity we might use to define the
dimensionless coefficient $\tilde K$ from $K$ and is given
by
\begin{eqnarray}
 N_c(g) =   2(4-D)(D-1)^2  \tr(1)^{-1} r_D  \Lambda_f^{2-D}
 \int_0^{\Lambda_p} dk  {k^{2D-5} \over M^2+r_D k^{D-2}},
\end{eqnarray}
which depends on two ultraviolet cutoff $\Lambda_f$ and
$\Lambda_p$.
The critical behavior of $\varphi$ near the critical $N_c$
is characterized by the critical exponent $\beta$ defined
by $\varphi \sim (N_c/N-1)^\beta$.  Hence the critical
exponent is given by
$\beta = 1/(D-2)$ for $2<D<4$.

\section{Chiral symmetry breaking and dynamical mass
generation}
\par
\subsection{critical coupling and phase diagram}
In what follows, for simplicity, we take the same cutoff
$\Lambda_f=\Lambda_f=\Lambda$.
Then, defining the dimensionless four-fermion coupling
constant $g$ by
\begin{eqnarray}
 g := M^{-2}\Lambda^{D-2} = G \Lambda^{D-2},
\end{eqnarray}
we obtain the critical number of flavors as a function of
$g$:
\begin{eqnarray}
 N_c(g) = N_c(\infty) [1-r_D^{-1} g^{-1} \ln (1+r_D g)],
 \label{critical}
\end{eqnarray}
where
\begin{eqnarray}
 N_c(\infty) := 2(4-D)(D-1)^2/[\tr(1)(D-2)],
\end{eqnarray}
and
\begin{eqnarray}
r_D  :=  {\tr(1) \Gamma(D/2)\Gamma(2-D/2) \over
4^{D-1}\pi^{(D-1)/2}\Gamma(1/2+D/2)} .
\end{eqnarray}
\par
The chiral symmetry is spontaneously broken for $N<N_c(g)$
where the critical number of flavors $N_c(g)$ depends on
the dimensionless four-fermion coupling constant $g$.
This implies the existence of the critical line
$N=N_c(g)$ in the phase diagram $(g,N)$.
The spontaneous chiral-symmetry breaking does not occur at
$g=0$, i.e., $N_c(0)=0$, as should does. The critical number
of flavors
$N_c(g)$ is monotonically increasing in $g$ and
remains finite in the whole range of $g$: $0 \le g \le
\infty$, i.e.,
$0 = N_c(0) \le N_c(g) \le N_c(\infty) < \infty $.
The massless vector boson limit $M \rightarrow 0$ (or the
limit of infinite four-fermion coupling constant $g
\rightarrow \infty$) can be taken in the arbitrary gauge
$\xi$ in this scheme. In particular, for
$D=3$, $N_c(\infty)=2$ ($r_3=1/8$).
\par
This result should be compared with QED3.
In QED3 the appearance of a critical $N_c$ was
shown in \cite{ANW88,KEIT95,KM95}, which
has been confirmed by the lattice Monte Carlo simulation
\cite{DKK89}. This issue was also analyzed in the scheme of
the effective potential in
\cite{SW92,SW89,Kondo95}.  The value $N_c(\infty)$
coincides with the critical number $N_c(\beta=0)$ in QED3
with the Lagrangian:
\begin{eqnarray}
{\cal L}_{QED3} = -{1 \over 4} \beta F_{\mu\nu}^2
+ \bar \psi^a i (\partial_\mu - ie A_\mu) \psi^a
+ {1 \over 2\xi} (\partial^\mu A_\mu)^2 ,
\end{eqnarray}
where the kinetic term for the gauge field vanishes in
the $\beta \rightarrow 0$ limit.
This coincidence is easily understood by comparing the
gauge boson propagator Eq.~(\ref{propagator}) with the
photon propagator in QED3 which is given by
\begin{eqnarray}
 D_{\mu\nu}^{(1)}(k)
 = {1 \over \beta k^2-\Pi_T^{(1)}(k^2)}
 \left( \delta_{\mu\nu}-{k_\mu k_\nu \over k^2} \right)
  +  {\xi \over k^2}{k_\mu k_\nu \over k^2},
\end{eqnarray}
since the gauge-parameter-dependent longitudinal part
does not contribute to the final result in the inversion
scheme. In the SD equation approach, it has been confirmed
\cite{KEIT95,KM95} that neglecting  higher powers
of the series in $k/\alpha$ does not
change the qualitative feature of the chiral phase
transition (infrared dominance).
\par
\subsection{dynamical generation of pole for
gauge boson}
\par
The gauge boson $A_\mu$ is merely the auxiliary field at
the tree level.   We have shown that for a given
$N<N_c(\infty)$ there is a critical value
$G=G_c(N)$ for the  four-fermion  coupling $G$ so that the
chiral symmetry is spontaneously broken  for $G>G_c(N)$,
which implies that the dynamical mass $m_f$ for the fermion
is generated for
$G>G_c(N)$, as long as $N<N_c(\infty)$.
The critical four-fermion coupling $G_c(N)$ is obtained by
solving Eq.~(\ref{critical}) with respect to $G$.
In the presence of the dynamical mass for the fermion, the
gauge boson propagator can have a pole mass $M_V$ (in the
time-like region) due to the massive fermion loop effect
even if the bare fermion mass is zero, $m=0$.
\cite{IKSY94}. The pole mass
$M_V$ is given by the equation
$G^{-1}=\Pi_T^{(1)}(-M_V^2;m_f)$ for $0 \le M_V^2 \le
4m_f^2$. Using the one-loop vacuum polarization tensor, the
ratio
$r_G:={M_V \over 2m_f}$ is given implicitly as the
solution of the equation \cite{IKSY94}:
\begin{eqnarray}
 m_f^{D-2} G
 = {3(4\pi)^{D/2} \over 4 \tr({\bf 1})\Gamma(2-{D \over 2})}
 \left[{}_2 F_1(2,2-{D \over 2},{5 \over 2}; r_G^2)r_G^2
\right]^{-1}.
 \label{poleeq}
\end{eqnarray}
The form of this equation is completely the same as that
derived already in \cite{Hands94}.  However the meaning of
this equation is conceptually different from each other.
Here this equation is used to search for the pole of the
dynamically generated {\it gauge} boson, not for the
auxiliary vector boson.
The ratio is monotonically decreasing function of $G$, and
goes to zero:
$r_G \sim 1/\sqrt{m_f^{D-2} G}$ as
$G \rightarrow \infty$ for arbitrary $D$, if it exists.
This implies that the fermion and antifermion is tightly
bound in the vector channel for strong
four-fermion coupling.
\par
For $2<D \le 3$, the solution
exists for any magnitude of the coupling constant $G$ as
long as $G>G_c(N)$, and $r_G$ in the small $G$ region is
given by
\begin{eqnarray}
r_G = 1 - {1 \over 2} \left( Gm_f^{D-2} {\Gamma({3-D \over
2}) \over 2^{D-2} \pi^{(D-1)/2}} \right) ^{2/(3-D)},
\end{eqnarray}
while in $D=3$
\begin{eqnarray}
r_G = 1 - \exp \left( -{2\pi \over m_f G} \right).
\end{eqnarray}
 For $3<D<4$, there exists a lower bound $G_V$ of
$G$ \cite{Hands94},
\begin{eqnarray}
G_V
 = {3(4\pi)^{D/2} \over 4 \tr({\bf 1})\Gamma(2-{D \over 2})}
 \left[{}_2 F_1(2,2-{D \over 2},{5 \over 2}; 1)
\right]^{-1} m_f^{2-D},
\end{eqnarray}
under which ($G<G_V$) the gauge boson propagator has no
pole and hence the bound state disappear
\cite{IKSY94}. Hence
a pole exists only when $G \ge max(G_c, G_V)$.
\par
\subsection{comparison with the SD equation}
\par
Our approach should be compared with the SD equation
approach done recently for this model \cite{IKSY94}
where $R_\xi$ gauge was used to
decouple the scalar mode $\theta$ from the theory after
formulating the Thirring model as a gauge theory by using
the hidden local symmetry. In the SD equation approach
the $R_\xi$ gauge quite simplifies the
formulation and the analysis of the SD equation as in our
analysis. However they have further introduced the nonlocal
version of the
$R_\xi$ gauge and taken a special nonlocal gauge in order
to eliminate the wavefunction renormalization for the
fermion, i.e., to guarantee $A(p) \equiv 1$ for the fermion
propagator
$S(p)=[{\dsl p}A(p)-B(p)]^{-1}$, since
the bare vertex was adopted to analyze the SD equation
for the fermion propagator and the bare vertex
approximation is justified, in light of the Ward-Takahashi
identity, only when there is no wavefunction
renormalization. This procedure greatly simplifies actual
analysis of the SD equation, since one has only to solve
the single integral equation for the fermion  mass function
$m(p):=B(p)/A(p)$.
\par
On the other hand, the non-local gauge leads to quite
complicated kernel for the integral equation of the mass
function.  The complexity prevents them from obtaining the
explicit solution and the explicit critical number of
flavors for general value of $G$.  Hence they have only
shown the existence of the nontrivial solution
corresponding to the bifurcation solution from the trivial
one $B(p) \equiv 0$, except for the special case $g=\infty$
at $D=3$ in which the explicit solution and the explicit
critical number of flavors can be obtained in completely
the same way as the QED3 \cite{KEIT95}.  They claim that
the phase transition associated with the spontaneous
breakdown of the chiral symmetry is the second order in the
sense that the phase transition is caused by the nontrivial
bifurcation solution besides the discontinuous one.
\par
Our approach has succeeded to derive almost all the
features on the chiral phase transition derived in the SD
equation approach. However we failed to show the scaling of
the essential singularity type, although we do not know
whether this type of scaling is correct or not for an
arbitrary $G$. We must say this problem is rather subtle in
our approach, see
\cite{KITE94}.  This point should deserve further studies.

\par
\subsection{technical remark}
\par
We here return to a technical problem.
In the expansion so far, the condition
$|{J \over 2k}|<1$ is assumed for an infinitesimal source
$J$.
We can show that the contribution from the region
$|{J \over 2k}|>1$ does not at all change the above result
as follows.  First note that
\begin{eqnarray}
&& D_{\mu\nu}^{(1)}(k) {\partial \over \partial J}
\Pi_{\mu\nu}^{(1)}(k)
\nonumber\\&=&
(D-1) M^{-2}
\left[1 - M^{-2} \Pi_T^{(1)}(k^2;J) \right]^{-1}
{\partial \over \partial J}
\Pi_T^{(1)}(k^2;J)
\nonumber\\&=&
(D-1)M^{-2}
\left( 1 + \sum_{p=1}^\infty
M^{-2p} [\Pi_T^{(1)}(k^2;J)]^p  \right)
{\partial \over \partial J} \Pi_T^{(1)}(k^2;J) .
\end{eqnarray}
If we perform the expansion
$\Pi_T^{(1)}(k^2;J) = \sum_{n=0}^\infty C_n
k^{2n+2}/J^{4-D+2n}$ with the $D$-dependent constant $C_n$,
then we have
\begin{eqnarray}
&&
(D-1)^{-1} \int_0^{2J} k^{D-1} dk
D_{\mu\nu}^{(1)}(k) {\partial \over \partial J}
\Pi_{\mu\nu}^{(1)}(k)
\nonumber\\
&=&
- \sum_{p=1}^\infty M^{-2p-2} \sum_{n_1} ... \sum_{n_p}
\sum_{l=0}^\infty C_{n_1} ... C_{n_p} C_l {(4-D+2l)
2^{D+4+2l+2\sum_{i=1}^p n_i}
\over D+4+2l+2\sum_{i=1}^p n_i}  J^{3D-5}
\nonumber\\ &&
- M^{-2} \sum_{l=0}^\infty C_l
{(4-D+2l)2^{D+2+2l} \over D+2+2l}
 J^{2D-3}.
\end{eqnarray}
Since $2D-3>D-1$ and $3D-5>D-1$ for $D>2$,
this contribution does not affect the above result when
$D>2$.

\section{Conclusion and discussion}
\par
In summary, we have given another reformulation of the
Thirring model as a gauge theory by introducing the
St\"uckelberg field as a Batalin-Fradkin field.  In this
standpoint the original Thirring model is identified with
the gauge-fixed version of a gauge theory and the equivalent
gauge theory has the well known BRS symmetry even after the
gauge-fixing.

\par
{}From the viewpoint of the effective potential, we have
shown the existence of the second order phase transition
(in the usual sense) associated with the spontaneous
breakdown of the chiral symmetry in the
$D$-dimensional Thirring model  $(2<D<4)$.   Up to the
leading order of
$1/N$ expansion,  the explicit critical number of flavors
$N_c$ was derived as a function of the four-fermion
coupling constant $G$ for arbitrary value of $G$, even for
$G=\infty$. All the above results are gauge-parameter
independent by construction.
\par
\par
Our approach based on the effective potential is also
extendable to analyze explicitly the non-Abelian case:
\begin{eqnarray}
 {\cal L} = \bar \psi^a i \gamma^\mu \partial_\mu \psi^a
 - {G \over 2N}(\bar \psi^a \gamma^\mu T^A \psi^a)^2,
 \label{nonabelianThirring}
\end{eqnarray}  with $T^A$ being the generator of a Lie
group $G$. In this case, the fictitious NG bosons
$\theta^A$ are not decoupled even in the $R_\xi$ gauge,
which makes the SD equation analysis rather complicated.
This model will be discussed in a separate paper.

\par
Finally we wish to point out that the Thirring model can be
identified with the gauged nonlinear sigma model.
Introducing the scalar field
$\varphi = \sqrt{{N \over 2G}} \exp (i \theta)$,
actually, the Lagrangian of the Thirring model is
rewritten into
\begin{eqnarray}
 {\cal L} = \bar \psi^a i \gamma^\mu (\partial_\mu
 - {i \over \sqrt{N}} A_\mu) \psi^a
 + |(\partial_\mu - {i \over \sqrt{N}}A_\mu)
\varphi|^2
 + {\cal L}_{\rm GF},
\end{eqnarray}
with a local constraint:
$
|\varphi(x)|^2 := \varphi^*(x) \varphi(x) = {N \over 2G}.
$
The analysis of the Thirring model as
the gauged nonlinear sigma model will be given elsewhere from the viewpoint of
the constrained system \cite{IKN95}.

\section*{Acknowledgments}
\par
The author would like to thank Atsushi Nakamura
for drawing his attention to the Batalin-Fradkin
formalism and Koichi Yamawaki for valuable discussions on
reference \cite{IKSY94}.
This work is supported in part by the Grant-in-Aid for
Scientific Research from the Ministry of Education, Science
and Culture (No.07640377).

\end{document}